\documentstyle[12pt]{article}
\begin{document}
\title{Analytic Structure of the Landau-Ginzburg Equation in $2+1$ 
Dimensions}
\author{Daniel Stubbs\thanks{E-mail address: ds@pineapple.apmaths.uwo.ca} \\ 
Department of Applied Mathematics \\ University of Western Ontario \\ 
London, ON \\ N6A - 5B7}
\maketitle

\begin{abstract}
In this paper, two methods are employed to investigate for which values of 
the parameters, if any, the two-dimensional real Landau-Ginzburg equation 
possesses the Painlev\'e property. For an ordinary differential equation to 
have the Painlev\'e property, all of its solutions must be meromorphic 
but for partial differential equations there are two inequivalent definitions, 
one a direct investigation of a Laurent series expansion in several complex 
variables and the other indirect and relying on the symmetry group of the 
partial differential equation. We check both methods for the Landau-Ginzburg 
equation in $2+1$ variables, and each one yields that this equation does not 
possess the Painlev\'e property for any values of the parameters. 
\end{abstract}

The Landau-Ginzburg equation arises in a wide variety of contexts in physics 
but the case for which we are interested is that of the order parameter of a 
binary fluid,
$$
\phi = \frac {c_1-c_2}{c_1+c_2}
$$
where $c_i$ represents that concentration of the $i$-th component of the fluid 
mixture. With the thermodynamic potential 
$$
L = \frac {1}{2}\left(-K|\nabla \phi|^2+a(T-T_c)\phi^2+\frac {b}{2}\phi^4
\right)
$$
we have that
\begin{eqnarray*}
\frac {\partial \phi}{\partial t} &=& \frac {\delta L}{\delta \phi} \\
                                  &=& -K\nabla^2\phi+a(T-T_c)\phi+b\phi^3
\end{eqnarray*}                                  
where $K>0$ is the surface tension between the two components of the fluid, 
$a>0$ and $b>0$ are system parameters and $T$ is the temperature of the fluid. 
Also, $T_c$ is a critical temperature at which the fluid undergoes a 
first-order phase transition.

There are two different definitions of the Painlev\'e property for a partial 
differential equation. The first definition is indirect, and relies upon a 
knowledge of the symmetry group of the partial differential equation: if $G$ 
is the group of continuous symmetries of a PDE, then this PDE has the 
Painlev\'e property if and only if every reduction of the PDE to an ODE by 
means of the group $G$ yields an ODE which possesses the Painlev\'e property.
It is more fully described in \cite{ar:olver}. The second definition, 
formulated in \cite{ar:weiss} and discussed in greater detail in \cite{bk:tabor}, 
is more direct: we expand the solution as a generalized Laurent series
$$
\phi = \Phi^{-\alpha} \sum_{j=0}^\infty \phi_j \Phi^j
$$
and then seek values of $\alpha$ and the $\phi_j$ such that this expansion is 
well-defined. In general, one has to place restrictions on the parameter 
values in order for such an expansion to exist. In each of these competing 
definitions, we are looking for an analytic crietria that will allow us to 
distinguish those equations whose solutions are, in some sense, ``regular'' 
and ``well-behaved''. The Painlev\'e property seeks to make this vague concept
of regularity precise by equating it with the property of a function being 
single-valued over a domain in complex $n$-space (where $n$ is the number of
independent variables), that is, of living on a one-sheeted Riemann surface. 

The direct method, particularly when the PDE has more than two independent 
variables, is often rather computationally intensive. In the case of the 
complex Landau-Ginzburg equation in one spatial variable, it was shown in 
\cite{ar:keefe} that only when the parameters yield the nonlinear 
Schr\"odinger equation does the Landau-Ginzburg equation have the Painlev\'e 
property. For two spatial dimensions, when we make our substitution 
we find that $\alpha=1$ and the resonance analysis reveals that there should 
be indeterminacies at $j=-1$ (corresponding to the indeterminacy of $\Phi$ 
itself) and at $j=4$. Working through the recursion relations yields
\begin{eqnarray*}
&\Phi^{-3}& \qquad \phi_0^2 = \frac {2K}{b}\left(\Phi_x^2+\Phi_y^2\right) \\
&\Phi^{-2}& \qquad 3b\phi_0^2\phi_1 = -\phi_0\Phi_t-K\left(2\Phi_x\frac 
{\partial\phi_0}{\partial x}+2\Phi_y\frac {\partial\phi_0}{\partial y}
+\phi_0\nabla^2\Phi\right) \\
&\Phi^{-1}& \qquad 3b\phi_0^2\phi_2 = \frac {\partial\phi_0}
{\partial t} - K\nabla^2\phi_0 -3b\phi_0\phi_1^2-a(T-T_c)\phi_0\\
&\Phi^{0} &\qquad 2b\phi_0^2\phi_3 = \frac {\partial \phi_1}{\partial
 t}+\phi_2\Phi_t+K\left(\nabla^2\phi_1+\phi_2\nabla^2
\Phi+
2\left[\frac {\partial \phi_2}{\partial x}\Phi_x+\frac {\partial \phi_2}
{\partial y}
\Phi_y\right]\right)- \\&& \hspace*{1.1in} b(\phi_1^3+6\phi_0\phi_1\phi_2)-
a(T-T_c)\phi_1 \\
&\Phi^{1}&\qquad 0\cdot \phi_4 = \frac {\partial\phi_2}{\partial t}+2\phi_3
\Phi_t
+K\left(
\nabla^2\phi_2+2\phi_3\nabla^2\Phi+4\left[\frac {\partial \phi_3}{\partial x}
\Phi_x + \frac {\partial \phi_3}{\partial y}\Phi_y\right]\right)-\\
 && \hspace*{1.1in} 3b(\phi_2^2
\phi_0+\phi_1^2\phi_2+2\phi_0\phi_1\phi_3)-a(T-T_c)\phi_2.
\end{eqnarray*} 
We should now use the first four equations of this set to successively 
eliminate $\phi_0$, $\phi_1$, $\phi_2$ and $\phi_3$ from the final equation. 
This equation will then contain only $\Phi$ and its various derivatives 
with respect to $x$, $y$ and $t$. Since $\Phi(x,y,t)$ is arbitrary, the 
coefficient of every independent term in that equation must vanish. This 
will give restrictions on the values of the parameters for which 
the generalized Laurent series is a valid expansion. In order to simplify 
these calculations, we often exploit a fundamental result from the theory 
of functions of several complex variables, specifically that none of the 
derivatives of $\Phi(x,y,t)$ vanish on the singular hypersurface $\Phi(x,y,t)
=0$. Hence we can use the implicit function theorem to eliminate one of the 
independent variables, say $t$, and re-write $\Phi$ as $t-\Psi(x,y)$. The 
coefficients $\phi_j$ then become functions only of $x$ and $y$, and our new 
set of recursion formul{\ae} become
\begin{eqnarray*}
 &\Phi^{-3}& \qquad \phi_0^2 = \frac {2K}{b}\left(\Psi_x^2+\Psi_y^2\right) \\
&\Phi^{-2}& \qquad 3b\phi_0^2\phi_1 = -\phi_0+K\left(2\Psi_x\frac 
{\partial\phi_0}{\partial x}+2\Psi_y\frac {\partial\phi_0}{\partial y}
+\phi_0\nabla^2\Psi\right) \\
&\Phi^{-1}& \qquad 3b\phi_0^2\phi_2 = - K\nabla^2\phi_0 -3b\phi_0\phi_1^2-
a(T-T_c)\phi_0\\
&\Phi^{0} &\qquad 2b\phi_0^2\phi_3 = \phi_2+K\left(\nabla^2\phi_1-\phi_2
\nabla^2\Psi-2\left[\frac {\partial \phi_2}{\partial x}\Psi_x+\frac 
{\partial \phi_2}{\partial y}\Psi_y\right]\right)- \\
&& \hspace*{1.1in} b(\phi_1^3+6\phi_0\phi_1\phi_2)-a(T-T_c)\phi_1 \\
&\Phi^{1}&\qquad 0\cdot \phi_4 = 2\phi_3+K\left(
\nabla^2\phi_2-2\phi_3\nabla^2\Psi-4\left[\frac {\partial \phi_3}{\partial x}
\Psi_x + \frac {\partial \phi_3}{\partial y}\Psi_y\right]\right)-\\
 && \hspace*{1.1in} 3b(\phi_2^2\phi_0+\phi_1^2\phi_2+2\phi_0\phi_1\phi_3)-
a(T-T_c)\phi_2.
\end{eqnarray*}
Successively solving these equations for the $\phi_i(x,y)$ yields
\begin{eqnarray*}
\phi_0 &=& \pm \sqrt{\frac {2K}{b}}\sqrt{\Psi_x^2+\Psi_y^2} \\
\phi_1 &=& \pm \frac {1}{3}\sqrt{\frac {K}{2b}}\left(\Psi_x^2+\Psi_y^2\right)
^{-1/2}\left[\nabla^2\Psi+2\left(\Psi_x\frac {\Psi_x\Psi_{xx}+\Psi_y\Psi_{yy}}
{\Psi_x^2+\Psi_y^2}+\right.\right.\\
& & \left.\left. \Psi_y\frac {\Psi_x\Psi_{xy}+\Psi_y\Psi_{yy}}{\Psi_x^2+
\Psi_y^2}\right)\right] \\
\phi_2 &=& \mp \frac {a(T-T_c)}{3\sqrt{2bK}} \mp \frac {1}{3}\sqrt{\frac 
{K}{2b}}\left(\Psi_x^2+\Psi_y^2\right)^{-1}\left\{ \nabla^2\sqrt{\Psi_x^2+
\Psi_y^2}+\frac {1}{6}\left(\Psi_x^2+\Psi_y^2\right)^{-1/2}\times\right. \\
& & \left. \left[\nabla^2\Psi + 2\left(\Psi_x\frac {\Psi_x\Psi_{xx}+
\Psi_y\Psi_{yy}}{\Psi_x^2+\Psi_y^2}+\Psi_y\frac {\Psi_x\Psi_{xy}+
\Psi_y\Psi_{yy}}{\Psi_x^2+\Psi_y^2}\right)\right]^2\right\}  
\end{eqnarray*}
The formul{\ae} obtained for $\phi_3$ and $\phi_4$ are extremely complex and 
not particularly illuminating; the final result is that there are {\em no} 
parameter values such that the expression for $\phi_4$ in terms of $\Psi(x,y)$
and its derivatives vanishes identically. Thus the original expansion lacks 
the required two arbitrary functions, since only $\Phi(x,y,t)$ is arbitrary 
but the general solution to a second-order PDE should possess two arbitrary 
functions. Hence the Landau-Ginzburg equation fails to have the Painlev\'e property for 
PDEs as defined by the direct method of \cite{ar:weiss}.

A basis of vector fields for the Lie algebra of the symmetry group of the 
two-dimensional Landau-Ginzburg equation can be calculated by means of a 
variety of symbolic algebra packages. Such a calculation yields the following 
list of generators:
\begin{eqnarray*}
{\bf v}_0 &=&\frac {\partial}{\partial t} \\
{\bf v}_i &=& \frac {\partial}{\partial x_i}, \quad i=1,2 \\
{\bf L} &=& x\frac {\partial}{\partial y} - y\frac {\partial}{\partial x} \\
{\bf D} &=& x\frac {\partial}{\partial x}+y\frac {\partial}{\partial y} +
 2t\frac {\partial}{\partial t}-\phi\frac {\partial}{\partial \phi}, \quad a=0
\end{eqnarray*}
This leads to the group $G =(E(2)\oplus \{ {\bf v}_0\}) \vdash \{ {\bf D} \}$, 
where $E(2)$ is the group of Euclidean motions in the plane. We must now 
determine what ordinary differential equations arise from using the above 
vector fields to reduce the Landau-Ginzburg PDE. In order to perform this 
reduction we first determine all subgroups of the complete symmetry group $G$ 
whose generic orbits have codimension one. From this knowledge we can then 
calculate the so-called symmetry variables $\eta(x,y,t)$ and $\rho(x,y,t)$, 
and then make the following transformation
$$
\phi = \rho(x,y,t)\tilde\phi(\eta(x,y,t))
$$
to obtain a first or second order ordinary differential equation. A more 
comprehensive account of this symmetry reduction procedure can be found in 
\cite{bk:olver}. This procedure leads to the following table:
$$
\begin{array}{cccc}
{\bf v}_0 &\eta =t & \rho=1 & a \ne 0\\
{\bf v}_i &\eta = x & \rho=1 & a\ne 0 \\
{\bf v}_0 \otimes {\bf v}_i &\eta = v(x+vt) & \rho =1 &a\ne 0\\ 
{\bf L} & \eta=r & \rho =1 & a \ne 0\\
{\bf L} & \eta =\theta &  \rho = 1/r & a=0\\
{\bf L} & \eta = -2B/(1+B^2)[\theta+B\log r] & \rho =2B/(r\sqrt{1+B^2}) &
 a=0 \\
{\bf D} & \eta = t/(2x^2) & \rho = 1/x & a=0  
\end{array} 
$$
Making the requisite transformations yields the following differential 
equations:
\begin{eqnarray*}
\frac {d\tilde\phi}{d\eta} - a(T-T_c)\tilde\phi - b\tilde\phi^3 &=& 0\\
K\frac {d^2\tilde\phi}{d\eta^2} - a(T-T_c)\tilde\phi-b\tilde\phi^3 &=& 0\\
K\frac {d^2\tilde\phi}{d\eta^2} + \frac {d\tilde\phi}{d\eta} -\frac {1}{v^2}
\left(a(T-T_c)\tilde\phi + b\tilde\phi^3\right) &=& 0\\
\frac {d^2\tilde\phi}{d\eta^2}+\frac {1}{\eta}\frac {d\tilde\phi}{d
\eta} - \frac {a(T-T_c)}{K}\tilde\phi -\frac {b}{K}\tilde\phi^3 &=& 0\\
\frac {d^2\tilde\phi}{d\eta^2}+\tilde\phi-\frac {b}{K}\tilde\phi^3 &=& 0 \\
\frac {d^2\tilde\phi}{d\eta^2}+\frac {d\tilde\phi}{d\eta} +\frac {B^2+1}
{4B^2}\tilde\phi - \frac {b}{K}\tilde\phi^3 &=& 0\\
\eta^2\frac {d^2\tilde\phi}{d\eta^2}+\left( 5\eta^2+\frac {1}{2}\right) \frac 
{d\tilde\phi}{d\eta}+2\tilde\phi-\frac {b}{K}\tilde\phi^3 &=& 0
\end{eqnarray*}
The first equation always possess the Painlev\'e property by virtue of a 
theorem of Fuchs that first-order ODEs which are analytic in both the 
dependent and independent variables only have poles of finite order as their 
movable singularities. The second and fifth equations in the list also possess
the Painlev\'e property, but the condition for the third equation to be 
Painlev\'e is that 
$$
a = -\frac {2v^2}{9K(T-T_c)}
$$
and so the equation cannot be Painlev\'e in a manner independent of the 
symmetry reduction. The fourth and seventh never have the property because of 
the presence of the independent variable in the coefficients and the sixth is 
Painlev\'e only when $B=\pm 3i$. Hence, when $a=0$ or when $a\ne 0$, not all 
of the ordinary differential equations possess the Painlev\'e property and so, 
by the indirect, group-theoretic criterion, the two-dimensional 
Landau-Ginzburg equation does not possess the Painlev\'e property for any 
values of the parameters.

Despite the failure of the Landau-Ginzburg equation to possess the Painlev\'e 
property in either the direct or indirect sense, we are still in a position 
to learn valuable information about the analytic structure of the 
Landau-Ginzburg equation. The seven ordinary differential equations listed 
above, and 
particularly the first four which are the only ones relevant for the case 
of $a\ne 0$, can all be subjected to further investigation in order to 
understand the precise manner in which they fail to be single-valued, and 
also to see if the multi-sheeted character of their Riemann surfaces 
eventually leads to the formation of a natural boundary in the complex 
plane beyond which solutions cannot be analytically continued. For instance, 
it is shown in \cite{ar:chang} that for the case of the Henon-Heiles system 
with non-integrable parameter values, the singularites arrange themselves 
into self-similar spirals with a fractal-like structure and so cause the 
formation of a natural boundary. One possible area of further investigation 
could then be to compare the numerical results obtained from integrating 
the above ODEs in the complex plane with the results obtained from the 
integration of the Landau-Ginzburg equation itself, using the spectral 
method discussed below. Another would be consist in allowing the system 
parameters, such as $a$, $b$ and $K$, to depend on either the spatial 
or temporal variables $x$, $y$ and $t$.

\end{document}